\documentstyle[12pt,fleqn,epsf]{ioplppt} 

\def \ba{\begin{array}}
\def \ea{\end{array}}

\def\be{\begin{equation}}
\def\ee{\end{equation}}
\def\beq{\begin{eqnarray}}
\def\eeq{\end{eqnarray}}

\def\nn{\nonumber}

\def\ve{\varepsilon}

\def\bc{\begin{center}}
\def\ec{\end{center}}
\def\bt{\begin{tabular}}
\def\et{\end{tabular}}

\begin{document}

\title{The problem of  singularities  caused by higher order curvature
corrections in four dimensional string gravity}

\author{M.Pomazanov${}^1$,
        V.Kolubasova${}^1$,
        S. Alexeyev${}^2$}

\address{${}^1$Department of Mathematics,
               Physics Faculty, \\
               Lomonosov Moscow State University,
               Moscow 119992, Russia}
\address{${}^2$Sternberg Astronomical Institute,
               Lomonosov Moscow State University, \\
               Universitetskii Prospect, 13,
               Moscow 119992, Russia}

\begin{abstract}
The  influence   of   higher   order   (stringly  inspired)  curvature
corrections to  the classical General Relativity spherically symmetric
solution is studied. In string  gravity  these  curvature  corrections
have a special  form  and can provide a  singular  contribution to the
field equations because they generate  higher  derivatives  of  metric
functions  multiplied   by   a   small   parameter.  Analytically  and
numerically  it  is shown  that  sometimes in  4D  string gravity  the
Schwarzschild  solution  is  not  recovered when the  string  coupling
constant  vanishes  and  limited  number  of  higher  order  curvature
corrections is considered.
\end{abstract}

\pacs{04.70.Dy, 04.20.Ex, 02.30.Hq, 02.60.Jh}

\section{Introduction}

The idea  of higher order  corrections to the Lagrangian of considered
system became a  rather common approach in modern theoretical physics.
Sometimes, these  additional corrections with higher order derivatives
can drastically modify (through a singular contribution) the solutions
of the  corresponding  Euler-Lagrange  equations.  This {\it singular}
contribution  does  not disappear, even when higher order  corrections
vanish. This fact  can  be illustrated with a  simple  example. Let us
consider the following  Lagrangian:
\begin{eqnarray}\label{example}
\tilde L = L(\dot q) + \ve^2 \ l(\ddot q),
\end{eqnarray}
where
\begin{eqnarray*}
L(\dot q) & = & \frac{\dot q^2}{2}, \\
l(\ddot q) & = & (-1)^k \ \frac{\ddot{q^2}}{2},
\end{eqnarray*}
$k=1, 2$. Corresponding Euler-Lagrange equations are:
\begin{eqnarray*}
\ddot q - (-1)^k \ \ve^2 \ q^{(4)} = 0.
\end{eqnarray*}
Let the initial conditions be
\begin{eqnarray*}
&& q(0) = q_o, \\
&& \dot{q}(0) =  v_o, \\
&& \ddot{q} (0) = 0, \\
&& q^{(3)} (0) = \delta q.
\end{eqnarray*}
As a  result,  one obtains two types of solutions. When $k = 1$ it has
the form
\begin{eqnarray*}
q = q_o + v_ot - \delta q \ \ve^3 \ \sin(t/\ve)
\end{eqnarray*}
and tends to the nonperturbed solution $q = q_o + v_o t$ when $\ve \to
0$. When $k=2$ it looks  like
\begin{eqnarray*}
q = q_o + v_o t + \delta q \ \ve^3 \ \mbox{sh} (t/\ve)
\end{eqnarray*}
and {\it has a singular contribution even when $\ve\to 0$}.

One meets  theories  with  higher  order  curvature corrections rather
often.  One  of the first examples is  the  wellknown  Wheeler-Feynman
electrodynamics  \cite{feynman}.  It is based on the  action  that  is
parametrized in the form of  two  dimensional  integral. Such integral
can be approximated by the  set  of one dimensional integrals, so,  an
infinite set  of  time  derivatives  appears  \cite{kerner}.  Modified
theories become unlocal  and  lead to  the  equations with the  higher
order derivatives up  to  the infinite  orders.  It is important  that
there is no problem  of  singular contribution from higher derivatives
because the solutions  are  limited and are infinitely differentiable.
Therefore,  the   lagrangian   of   infinite  order  approximates  the
lagrangian of second  order which provides the regular contribution of
higher order corrections.

We would like  to point  out that  the  problems with  new degrees  of
freedom, and, so,  appearing  of nonphysical ``runaway'' solutions are
really actual for  one  special  class of the models.  There  are  the
models  where  the  corrections  contain  high  but  finite  order  of
derivatives \cite{bento},  where  the  singular contribution of higher
order  correction  is not compensated by the  contribution  from  next
order(s).  The  most natural approach to  such  class of models is  to
introduce additional limits on solutions availability  and to consider
only such  ones that have a regular part  (expanding in Taylor series)
relatively the  numerical  parameter  before  the  higher  derivatives
\cite{simon}. If one considers  such  condition as necessary one there
would be no problem of higher order corrections  (In Ref. \cite{simon}
the problem is treated in such a manner).

Here  it  is   necessary  to  emphasis  that  higher  order  curvature
corrections do  not  automatically  cause singular unlimited solutions
(see the example higher, Eq. (\ref{example})). So, it  is desirable to
study the conditions of appearance  of  new  ``runaway solutions'' and
their status  in the four  dimensional (4d) gravity models with higher
curvature corrections caused by string  theory  more  carefully. It is
important to extract the properties of the solutions (in this paper we
restrict our consideration with spherically symmetric  ones) when only
limited (finite) number of corrections are considered. Further, during
consideration of real physical problems it would be better to restrict
consideration with models where higher  order  corrections  (that  are
necessary to make a little step from General  Relativity boundaries of
applicability) do not  cause singular contribution to the regular part
of  the solution.  The  stability of the  solution  under the  initial
conditions  on  the infinity is strongly required  from  the  physical
grounds.

In the perturbational approach, the  4d  effective  (string  inspired)
gravity action has the following form (we work in Planckian  system of
units where $\hbar = c = m_{Pl} = 1$)
\cite{c01}:
\begin{eqnarray}\label{act1}
S &  = & \frac{1}{16\pi} \int d^4 x \sqrt{-g} \biggl[ - m^2_{Pl} R + 2
\partial_{\mu} \phi \partial^{\mu} \phi \nonumber \\
& + & \lambda e^{-2 \phi}  l_2 + \lambda^2 e^{-4 \phi} l_3 + \lambda^3
e^{-6 \phi} l_4 + O(\lambda^4) \biggr],
\end{eqnarray}
where $R$ is the Ricci scalar, $\phi$ is a dilatonic  field, $\lambda$
is  a  string  coupling  constant  (an  expansion  parameter  which is
proportional  to   string   tension   $\alpha'$   with  the  numerical
coefficient  depending  upon  the  type of string gravity)  and  $l_i$
($i=2, 3, \ldots$) are higher order  curvature  corrections.  In  this
paper we study the class  of  the models where higher order  curvature
corrections  consist from  Riemannian  tensor  $R_{\mu\nu\alpha\beta}$
products. Such type of models are widely discussed  now from different
aspects  \cite{c07}.   This   study  focuses  on  asymptotically  flat
spherically symmetric static black hole solutions.

It should also be  pointed out that the zeroth measure of  the initial
conditions manifold (only asymptotically flat,  for  example)  is  not
very realistic, especially when new observational results in cosmology
are taken into  account.  The highly  probable  existence of a  global
nonvanishing cosmological  constant  $\Lambda$  requires the extension
for the manifold of initial conditions. One can  neglect the $\Lambda$
influence  on the  global  structure of the  solution  because it  has
zeroth order  contribution  relatively  higher  derivatives,  so, it's
influence  is  important   only  in  the  regions  with  rather  small
curvature.  In  black  hole  topologies   where   its   influence   is
neglectible,  it  is  possible  to restrict the consideration  by  the
requirement  of  stability  (of  Lyapunov  type)   under  the  initial
conditions (and, hence, to  take  $\Lambda$ existence into account) at
the infinity.

The most convenient choice of four dimensional metric is, therefore,
\begin{equation}\label{act2}
ds^2 =  - \Delta(r) dt^2 + \frac{\sigma ^2(r)}{\Delta  (r)} dr^2 + r^2
\biggl( d \theta^2 + \sin^2 \theta d \varphi^2 \biggr).
\end{equation}
The solution  of  corresponding  nonperturbed  field  equations is the
wellknown Schwarzschild one
\begin{eqnarray}\label{act3}
\Delta & = & \Delta^0 (r) = 1 - \frac{2M}{r}, \nonumber \\
\sigma & = & \sigma^0 (r) = 1, \\
\phi & = & \phi^0 (r) = \phi_0 = \mbox{const}, \nonumber
\end{eqnarray}
where $M$ is black hole mass.

When the correction $l_2 = R_{ijkl}  R^{ijkl} - 4 R_{ij} R^{ij} + R^2$
in  (\ref{act1}) is  taken  into account (which  is  the wellknown  4d
curvature invariant  named  as  Gauss-Bonnet  term), the corresponding
solutions  do  not  contain  higher  derivatives.  It is necessary  to
emphasize that the discussed black  hole  solution  behavior  strongly
differs from usual Schwarzschild one under the event horizon as it was
shown in  detail in Ref. \cite{c02}.  When the metric  (\ref{act2}) is
substituted in (\ref{act1}), the action becomes \cite{c03}
\begin{eqnarray}
S  =  \int  dr  \biggl[ L_\Upsilon(r, \Delta, \sigma,  \phi,  \Delta',
\phi',  \lambda)  + \lambda ^2 l_\Upsilon (r,  \Delta,  \sigma,  \phi,
\Delta', \sigma', \Delta'', \lambda) \biggr]
\end{eqnarray}
where  $\Upsilon  = b, h, s$  corresponds  to the three string  theory
types  considered  in  this  paper:  bosonic  (b),  heterotic  (h) and
superstring II (s).

The basic Lagrangian does not contain higher derivatives,
\begin{eqnarray}\label{act4}
L_b & = & L_h = L_0(r, \Delta, \sigma, \phi, \Delta', \phi') + \lambda
L_{GB} (r,\Delta, \sigma, \phi, \Delta', \phi'), \nonumber \\
L_s & = & L_0 = - \frac{\Delta' r  + \Delta  + \sigma^2  - \Delta  r^2
(\phi')^2} {\sigma}, \\
L_{GB}  & =  &  4 e^{-2\phi} \phi'  \Delta'  \frac{\Delta (\phi')^2  -
\sigma^2} {\sigma^3}. \nonumber
\end{eqnarray}
$L_0$ is simply the contribution  resulting from General Relativity.

The  next  order  curvature  corrections  to   $l_\Upsilon$  have  the
following form \cite{c04}:
\begin{eqnarray}\label{act5}
l_b & = & C_1 \sqrt{-g} e^{-4\phi }(
2 R^{\mu\nu }{}_{\alpha\beta }R^{\alpha\beta }{}_{\gamma\rho }
R^{\gamma\rho }{}_{\mu\nu }
-4R^{\mu\nu }{}_{\alpha\beta }R_{\nu }{}^{\gamma\beta\rho }
R^{\alpha}{}_{\rho\mu\gamma }  \nonumber \\
& + & \frac{3}{2} R R^2_{\mu\nu\alpha\beta }
+ 12 R^{\mu\nu\alpha\beta }R_{\alpha\mu }R_{\beta\nu }
+8 R^{\mu\nu }R_{\nu\alpha }R^{\alpha }{}_{\mu }
- 12 R R^2_{\alpha\beta } + \frac{1}{2} R^3 \nonumber \\
& + & R^{\mu\nu }{}_{\alpha\beta }R^{\alpha\beta }{}_{\gamma\rho }
R^{\gamma\rho }{}_{\mu\nu }) + O(\lambda), \nonumber \\
l_h & = & C_2 \sqrt{-g} e^{-6\phi }( A
-\frac{1}{8}(R_{\mu\nu\alpha\beta }R^{\mu\nu\alpha\beta })^2
-\frac{1}{4}R_{\mu\nu }{}^{\gamma\delta }R_{\gamma\delta }
{}^{\rho\eta }
R_{\rho\eta }{}^{\alpha\beta }R_{\alpha\beta }{}^{\mu\nu }
\nonumber \\
& + & \frac{1}{2}R_{\mu\nu }
{}^{\alpha\beta }R_{\alpha\beta }{}^{\rho\eta }
R^{\mu }{}_{\eta\gamma\delta }R_{\rho }{}^{\nu\gamma\delta }
+ R_{\mu\nu }{}^{\alpha\beta }R_{\alpha\beta }{}^{\rho\nu }
R_{\rho\eta }{}^{\gamma\delta }R_{\gamma\delta }{}^{\mu\eta })
+ O(\lambda), \\
l_s & = & C_3 \sqrt{-g} e^{-6\phi } A + O(\lambda), \nonumber
\end{eqnarray}
where
\begin{eqnarray*}
A  =  \zeta(3) [R_{\mu\nu\rho\eta}  R^{\alpha\nu\rho\beta} (R^{\mu\nu}
{}_{\delta\beta}  R_{\alpha\gamma}  {}^{\delta\eta}   -  2  R^{\mu\nu}
{}_{\delta\alpha} R_{\beta\gamma} {}^{\delta\eta} )],
\end{eqnarray*}
$R^2_{\mu\nu}  =  R_{\mu\nu}  R^{\mu\nu}$,  $\zeta(3)$  is   Riemanian
zeta-function, $C_{1,2,3}$ are numerical coefficients.

The influence of higher order curvature corrections to the behavior of
spherically symmetric  static  solutions  was considered in \cite{c03}
where a perturbed solution close the nonperturbed one  was studied. It
was obtained by a specially  developed  and  coded numerical iteration
method. The only indefinite  point  of the perturbed solution occurred
near the particular point $r_s$ (internal black hole  singularity of a
new type at finite distance from the origin,  provided by Gauss-Bonnet
term, see  Ref. \cite{c02}). If  one works  in the frames  of such  an
approach,  the   singular   contribution   disappears.   From  a  pure
mathematical  point  of view, the solution tends  to  some  particular
branch that begins from a  particular  manifold  of initial conditions
\cite{levin}  with  null  measure.  Even  small   changes  of  initial
conditions cause an appearance of  a  singular  contribution that does
not vanish when $\ve\to 0$.

Finally, the main  aim  of  this paper is to  understand  under  which
circumstances  a  singular contribution is induced in three  different
closed 4d low energy string gravity models on the Schwarzschild metric
background when  radial coordinate $r$  is rather large, when the only
{\it finite} number of higher  order  curvature  corrections are taken
into account (as  the  complete expansion  is  unknown). The paper  is
organized as  follows: In Section  2 we show some general mathematical
considerations of  higher  order  singular  corrections  behavior,  in
Section 3 we apply these results to bosonic, heterotic and superstring
II (SUSY II) closed 4d low  energy models, Section 4 contains the SUSY
II numerical investigations, Section 5 is  discussions and conclusions
one.

\section{General theory of singular contributions}

Models with  corrections  including  higher  order  derivatives can be
generically written as:
\begin{eqnarray}\label{act6}
\tilde L (t,x,\dot x, \ddot x) = L(t,x,\dot x ) + \ve^2 l(t,x,\dot{x},
\ddot{x}),
\end{eqnarray}
where $\ve$ is an expansion parameter, $t$ is real variable, $x(t)$ is
a smooth  vector of $n$-dimensional manifold  and $L$, $l$  are smooth
functions.

The corresponding Euler-Lagrange equations are
\begin{eqnarray*}
\frac {\partial {\tilde L}}{\partial x} - \frac {d}{dt} \frac{\partial
{\tilde L}} {\partial {\dot x}}  +  \frac  {d^2}{dt^2} \frac {\partial
{\tilde L}} {\partial {\ddot x}} = 0.
\end{eqnarray*}
One can rewrite these equations as follows
\begin{equation}
\label{act7}
\begin{array}{l}
{\displaystyle  \ve^2 x^{(4)}  \frac{\partial^2  l}{\partial  \ddot{x}
\partial \ddot{x}}  =  \ddot{x}  \frac{\partial^2  L}{\partial \dot{x}
\partial \dot{x}} + \Phi (t, x, \dot{x}) + } \\ \\
{\displaystyle +  \ve^2  \left[-x^{(3)}  \frac{\partial^3 l} {\partial
\ddot{x}  \partial  \ddot{x} \partial \ddot{x}} x^{(3)} + x^{(3)}  \Xi
(t, x, \dot{x}, \ddot{x}) + \Psi (t, x, \dot{x}, \ddot{x})\right],}
\end{array}
\end{equation}
where  $\displaystyle\frac{\partial^2   L}{\partial  \dot{x}  \partial
\dot{x}}$  and   $\displaystyle\frac{\partial^2  l}{\partial  \ddot{x}
\partial \ddot{x}}$ are symmetric matrixes of second order derivatives
of $L$ and  $l$ (respectively) whereas  $\Phi$, $\Xi$ and  $\Psi$  are
vector  matrixes with  second  and third derivatives  of  $L$ and  $l$
(respectively).

Introducing  the  following  change  of  variables:   $\dot{x}  =  y$,
$\ddot{x}  =  z$,  $\ve  x^{(3)}  =  v$  and  considering  the  matrix
$\frac{\partial^2l}{\partial   \ddot{x}   \partial  \ddot{x}}$   as  a
nondegenerated one, we can write (\ref{act7}) in the form
\begin{equation}
\label{act8}
\left\{
\begin{array}{l}
\ve\dot v= F(t, x, y, z, \ve), ~~ \ve\dot z=v, \\
\dot y=z,~~ \dot x=y.
\end{array}
\right.
\end{equation}
where
\begin{eqnarray}
F(t, x, y, z, \ve) & = &  \left[\frac{\partial^2l}{\partial z \partial
z} (t,x,y,z)  \right]^{-1}  z  \frac{\partial^2 L}{\partial y \partial
y}(t,x,y)   +   \left[\frac{\partial^2   l}{\partial  z  \partial   z}
\right]^{-1} \Phi (t,x,y) \nonumber \\
& - &  \left[\frac{\partial^2 l}{\partial z \partial z} \right]^{-1} v
\frac{\partial^3  l}{\partial  z  \partial  z  \partial  z}  v  +  \ve
\left[\frac{\partial^2  l}{\partial  z  \partial  z}  \right]^{-1}   v
\Xi(t,x,y,z) \nonumber \\
&  +  &   \ve^2   \left[\frac{\partial^2  l}{\partial  z  \partial  z}
\right]^{-1} \Psi (t, x, y, z). \nonumber
\end{eqnarray}

When  $\ve=0$  the  system  (\ref{act8})  has  a  degenerated  solution
$\stackrel{0}{X}$,
\begin{equation}
\begin{array}{l}\label{act9}
x = x^0(t),\ y=y^0(t),\ z=z^0(t),\ v=v^0(t)\equiv 0,
\end{array}
\end{equation}
which corresponds to the nonperturbed one  of Euler-Lagrange equations
(\ref{act6})   when   the   initial  conditions  are   $x^0(t_0)=x_0$,
$y^0(t_0)=y_0$. This means  that  the equations (\ref{act8}) belong to
A.N.Tikhonov's  class  \cite{tikhonov}.  According   to   the  related
theorem, the closeness  of perturbed and nonperturbed solutions can be
obtained, with initial conditions in the neighborhood of $x_0$, $y_0$,
$x''(t_0)$, $x'''(t_0)$, if all the real parts of eigenvalues $\omega$
of matrix
$$\left(
\begin{array}{cc}
0 & {\displaystyle\frac{\partial F}{\partial z}}\\ & \\
I & 0
\end{array}
\right)
$$
are negative.
This matrix  is calculated on nonperturbed solutions $\stackrel{0}{X}$
(\ref{act9}), $I$ is identity $n \times n$ matrix.

After  some  algebra,  the  equation  on  eigenvalues $\omega$ can  be
represented as
$$
\mbox{det} \left|\omega^2 \frac{\partial^2 l}{\partial z \partial z} -
\frac{\partial^2 L}{\partial y \partial y} \right|_{\stackrel{0}{X}} =
0.
$$
Therefore, to provide the convergence of the perturbed solution to the
nonperturbed one, it  is possible to  require only $Re(\omega)  =  0$.
This corresponds to the case when the value
\begin{equation}
\label{act10}
u(t)\leq 0,
\end{equation}
where $u(t)$ can be calculated from the following equation
\begin{equation}
\label{act11}
\mbox{det} \left|u(t)  \frac{\partial^2 l}{\partial \ddot{x}  \partial
\ddot{x}}  -  \frac{\partial^2  L}{\partial \dot{x} \partial  \dot{x}}
\right|_{X = \stackrel{0}{X}(t)} = 0.
\end{equation}
We shall  call $u(t)$ the {\it  singular index}. Eq.  (\ref{act11}) is
valid  on  the  interval  $t\in[t_0,   T]$   where   the  solution  of
nonperturbed problem exists when $\ve=0$.

It is  important to emphasize  that the condition (\ref{act10}) is not
enough to ensure  the  absence of  any  singular contribution to  Eqs.
(\ref{act6}),  but  when  the  condition  (\ref{act10})  is  broken  a
singular  contribution  of  order  $\displaystyle  \mbox{exp}  \biggl(
\frac{t}{\ve} \ \mbox{Re} \sqrt{u(t)} \biggr)$ appears necessarily.

Switching to the string gravity perturbed Lagrangian
\begin{eqnarray*}
\tilde L = L(r,\Delta, \sigma,  \phi,  \Delta', \phi') + \ve^2 \  l(r,
\Delta, \sigma, \phi, \Delta', \sigma', \Delta'')
\end{eqnarray*}
and taking  into account that  $\phi(r)$ is not perturbed,
one can conclude  that the condition  (\ref{act10})  on
$u(r)$ can be fulfilled if
\begin{equation}
\label{act12}
\left|u\,P - Q\right|_{\Delta = \Delta^0(r), \ \sigma = \sigma^0(r),\
\phi = \phi^0(r)} = 0,
\end{equation}
where
$$P = \left(
\begin{array}{cc}
l_{\Delta ''\Delta ''} & l_{\sigma '\Delta ''}\\
l_{\Delta ''\sigma '} & l_{\sigma '\sigma '}
\end{array}
\right)$$
and
$$Q=\left(
\begin{array}{ccc}
L_{\Delta'  \Delta'}   -  L_{\Delta'  \phi'}  \left[L_{\phi'  \phi'  }
\right]^{-1} L_{\phi' \Delta'}
& &
L_{\sigma\Delta'}   -    L_{\sigma   \phi'}   \left[L_{\phi'    \phi'}
\right]^{-1} L_{\phi'  \Delta'}  \\  L_{\Delta'  \sigma}  - L_{\Delta'
\phi'} \left[L_{\phi' \phi'} \right]^{-1} L_{\phi' \sigma}
& &
L_{\sigma\sigma } -L_{\sigma \phi'} \left[L_{\phi' \phi'} \right]^{-1}
L_{\phi' \sigma}
\end{array}
\right)$$

\section{Singular indexes of low energy effective string gravity}

To investigate the  problem of singular contributions in string theory
Lagrangians with  higher  order  curvature  corrections in spherically
symmetric space times ,it is necessary,  as a first step, to apply the
formulas (\ref{act4}) and (\ref{act5}) to determine  the accurate form
of the main Lagrangian and higher  order  curvature  corrections.  The
software packages from MAPLE  and  REDUCE were specially developed and
used to  obtain the corresponding singular  indexes. With the  help of
Eq. (\ref{act12}) and  after  the substitution of Schwarzschild values
of  metric  functions  (\ref{act3})  in  the  limit  $r/M  \rightarrow
\infty$, we get the following asymptotic formulas.

For bosonic case
$$
u^b_{1, 2} = \frac{ e^{4 \phi_0} r^2 (1 \pm i \sqrt{15})} {576} \left[
\frac{r}{M} + o \left(\frac{r}{M} \right) \right].
$$

For heterotic case
$$
\begin{array}{l}
{\displaystyle
u^h_1=0.008773e^{6\phi _0}r^4
\left[\frac{r^2}{M^2}+o\left(\frac{r^2}{M^2}\right)\right],\
} \\ {\displaystyle
u^h_2=-0.02291e^{6\phi _0}r^4
\left[\frac{r^2}{M^2}+o\left(\frac{r^2}{M^2}\right)\right]
}
\end{array}
$$

For SUSY II case
$$
\begin{array}{l}
{\displaystyle
u^s_1=-0.03496e^{6\phi _0}r^4
\left[\frac{r^2}{M^2}+o\left(\frac{r^2}{M^2}\right)\right],\
} \\ {\displaystyle
u^s_2=-0.198e^{6\phi _0}r^4
\left[\frac{r^2}{M^2}+o\left(\frac{r^2}{M^2}\right)\right]
}
\end{array}\nn
$$

Considering  these  asymptotical formulae, one can conclude that  only
SUSY II  does not induce any singular contribution  to the solution in
the neighborhood of  the  Schwarzschild  one , even for  big  $r$.  To
understand the perturbed  solution  behavior in  SUSY  II case, it  is
necessary  to  perform a  numerical  investigation  of  this  solution
dependence upon  different sets of initial conditions $\Delta''(r_0)$,
$\Delta'''(r_0)$, $\sigma'(r_0)$, $\sigma''(r_0)$, that  are  close to
the corresponding Schwarzschild values $\Delta''_0 (r) = -4  M / r^3$,
$\Delta'''_0 (r) = 12M / r^4 $, $\sigma'_0 (r) = \sigma'''_0 (r) = 0$.

\section{Results of  numerical investigation of perturbed solutions in
SUSY II}

To investigate the difference  between  the perturbed SUSY II solution
and the nonperturbed Schwarzschild  one  , we solved Eqs. (\ref{act7})
as the Cauchy problem starting from initial point  $r_0$ using initial
conditions  from   accurate   Schwarzschild   solution.  A  7th  order
Runge-Kutta  code  was especially chosen for the  integration  of  the
resulting system. Some computing problems occurred in the range of big
$r_0/M$. For instance, when $\lambda = 0.1$, $M = 1$ for $r_0 > 6$ (we
work in Planckian units)  the system becomes so rigid that it  can not
be solved.  We could integrate these equations only  in the range that
was situated up to the event horizon at $r/M \in [2.1,5.2]$ and for $M
< 100$,  because for higher mass values and  higher $r/M$ values local
influence of the corrections became so small that calculations stopped
at the first integration step.

For  different  values  of  $r_0$,  the   difference  of  $\Delta(r)$,
$\sigma(r)$ and $\phi(r)$ from corresponding Schwarzschild values with
high accuracy (not less than integration method one) were proportional
to $(r-r_0)^2$, so,
\begin{equation}\label{act13}
\delta x = C (r - r_0)^2,
\end{equation}
where $C$ is numerical coefficient.

We investigated the  dependence  of $C$ as a  function  of the initial
integration point $r_0$ and mass $M$ when $\lambda$ is fixed and equal
to $0.1$\footnote{The increasing of $M$ when  $\lambda$  is  fixed  is
equivalent (up to  dimensionality) to decreasing of $\lambda$ when $M$
is  fixed}. So, the  behavior  of  $C(r_0,  M)$ is  described  by  the
following fit
$$
C(r_0, M) = \alpha(M) \biggl( \frac{r_0}{M} \biggr)^{-\beta(M)}.
$$
It  is  important  to  emphasize  that  $\beta(M)>10$  has  very  weak
dependence  upon  $M$.   For   instance,  the  results  for  $\phi(r)$
($\phi_0(r) \neq 0$) are presented in the Table 1.

\begin{table}
\centerline{\begin{tabular}{|c|c|c|}\hline
$M$ & $\alpha$             &  $\beta$ \\ \hline \hline
1   &  134.                &  13.53   \\ \hline
2   &  0.411               &  13.59   \\ \hline
15  & $4.39\cdot 10^{-8}$  &  13.81   \\ \hline
50  & $3.29\cdot 10^{-12}$ &  13.81   \\ \hline
100 & $1.49\cdot 10^{-14}$ &  13.92   \\ \hline
\end{tabular}}
\caption{The  dependence  of the numerical fit coefficients of  $\phi$
expansion $\alpha$ and $\beta$ versus mass $M$}
\end{table}

If one extrapolates  formula  (\ref{act13}) for $\delta x$ (difference
between perturbed SUSY II solution and nonperturbed Schwarzschild one)
in the complete range of  $r/M  \in (2, \infty)$, after obtaining  the
asymptotically flat solution (starting from  $r_0  \to  \infty$),  one
would conclude (for the fixed $r$ and $\beta$ from Table 1) that:
$$
\lim_{r_0  \to  \infty}  C(r_0,M)  (r-r_0)^2 = \lim_{r_0  \to  \infty}
C(r_0, M) r_0^2 \Biggl( 1 + o \biggl( \frac{r}{r_0} \biggr)  \Biggr) =
0.
$$
Then it is possible to estimate $\delta x (r)$ as
$$
\delta x(r) < \max_{r_0 \in [r,\infty)} C(r_0, M)  (r-r_0)^2 = \frac{4
\alpha (M)}{(\beta  -  2)^2} \left(\frac{\beta}{\beta - 2} \frac{r}{M}
\right)^{-\beta} r^2.
$$
For example, at the event horizon $r = r_h = 2M$ :
$$
\delta x (2M) < D(\beta) \alpha (M) M^2,
$$
where
$$
D (\beta  ) = \frac{16} {(\beta - 2)^2}  \left( \frac{2 \beta}{\beta -
2} \right)^{-\beta}.
$$
If the value  of  parameter $\beta$  does  not differ apparently  from
those  presented  in  Table  1  and  the dependence  $\alpha  (M)$  is
approximately the same, one can conclude that
$$
\lim_{M \rightarrow \infty} \alpha (M) M^2 = 0.
$$
So, the value of the perturbation becomes zero when relative weight of
higher order corrections vanishes.

Finally, this, as it seems to the authors of the paper, means that the
4th  order  curvature correction  in  SUSY  II  does  not  provide any
appreciable differences from  the  Schwarzschild solution, at least up
to the event horizon.

\section{Discussion and conclusions}

Based  on our  preliminary  results in the  framework  of models  with
higher  order  curvature  corrections  made of pure  Riemanian  tensor
products, it is reasonable to assume that the only model that does not
provide any singular  contribution is the  SUSY II one.  As  curvature
corrections of next orders do not produce new higher derivatives, in a
case  if  singular contributions are absent in  the  4th  perturbation
order,  they  would  be  absent  in  all perturbative  orders.  It  is
impossible to draw more accurate  conclusions  in  this framework. Our
results show the existence of  singular  contributions  in bosonic and
heterotic realizations  of string gravity and  give hope that  SUSY II
model  can  be  free   from   singular  contributions,  at  least  for
spherically symmetric static space times.

It is necessary to emphasis that the loop expansion of  string gravity
(the origin of higher order curvature  corrections) initially consists
from  the  terms with increasing order of derivatives  \cite{c04,c05}.
But working  in the frames only  of one considered  approximation (for
example, 3rd, 4th,  ...)  it is  possible  to make some  manipulations
\cite{c04,c05} taking usage from the fact that corrections contain the
products with dilaton.  So, analogously to the transformation that was
used when  the Gauss-Bonnet term by itself was  changed to its current
\cite{c02},  one  can  make  partial  integration  of  the  correction
according to the scheme:
\begin{eqnarray}\label{conc}
\int e^{n \Phi} \nabla_n R^n =
(e^{n \Phi} \nabla_{n-1} R^{n})
- \int n \nabla\Phi e^{n \Phi} \nabla_{n-1} R^n.
\end{eqnarray}
Making this procedure ``n''  times  one can shift differentiation from
Riemannian tensor to  the dilaton and  finally result with  the  terms
like $R_{ijkl}^n$ without derivatives. Then, as one is  working in the
limits of corresponding correction, the term  with dilaton derivatives
can be  transferred  to a next order as it is shown in Formula (10) at
Ref.  \cite{c06}.  Hence, {\it  in  the  frames  of  chosen  order the
expression becomes free from derivatives}.

So, the result of our study  is that when the limited number of higher
order  curvature  corrections  are  taken into account  a  spherically
symmetric solution  can  disappear. Finally, the singular contribution
of the  higher order curvature corrections in the  low energy limit of
string  gravity,  in  the  frames of Lagrange approach,  represents  a
problem. Anyway, as the  consideration  of higher order corrections is
necessary  to  extend  the  boundaries of applicability  of  classical
gravity (and to obtain preliminary  directions  on  some effects, that
have quantum gravity nature) this can be avoided by two possible ways:
\begin{itemize}
\item
to suggest  additional  principles  for  choosing  the special initial
conditions for the  higher derivatives, providing the closeness of the
perturbed solution to the nonperturbed one (at least for big $r$) that
include all  the  physically  interesting  cases (asymptotically flat,
dS/AdS);
\item
not to restrict the consideration by the frames of chosen perturbative
order and  consider the ``rest  term'' of series expansion (this means
that the application of Formula (10) from \cite{c06} and all  the same
ones are not directly allowed in black hole cases).
\end{itemize}

\section*{Acknowledgments}

S.A.  would like to  thank  the  AMS  Group in  the  ``Laboratoire  de
Physique Subatomique et de Cosmologie  (CNRS/UJF)  de  Grenoble''  for
kind hospitality. This work was supported in part by ``Universities of
Russia: Fundamental Investigations'' via grant No. UR.02.01.026 and by
Russian Federation State Contract No. 40.022.1.1.1106. The authors are
grateful to  A. Barrau and G. Boudoul for  the very useful discussions
on the subject of this paper.


\section*{References}

\begin{thebibliography}{99}

\bibitem{feynman}
J.A.Wheeler  and  R.P.Feynman,  {\it  Rev. Mod. Phys.} {\bf  21},  425
(1949).

\bibitem{kerner}
E.J.Kerner, {\it J. Math. Phys.} {\bf 3}, 35 (1962).

\bibitem{bento}
M.C. Bento,  O. Bertolami, {\it Phys.  Lett.} {\bf B228},  348 (1989),
{\bf B368}, 198 (1996).

\bibitem{simon}
J.Z.Simon, {\it Phys. Rev.} {\bf D 41}, 12 (1990).

\bibitem{c01}
A.Tseytlin,
``String Solutions with Nonconstant Scalar Fields''  {\it Published in
the  proceedings   of  International  Symposium  on  Particle  Theory,
Wendisch-Rietz,     Germany,     7-11     Sep     1993     (Ahrenshoop
Symp.1993:0001-13),} hep-th/9402082;

B.Zwiebach, {\it Phys.Lett.} {\bf B156}, 315  (1985);

E.Poisson, {\it Class.Quant.Grav.} {\bf 8}, 639 (1991);

D.Witt, {\it Phys.Rev.} {\bf D38}, 3000 (1988);

J.T.Wheeler, {\it Nucl.Phys.}  {\bf B268}, 737 (1986), {\bf B273}, 732
(1986);

G.W.Gibbons and K.Maeda, {\it Nucl.Phys.} {\bf B298}, 741 (1988);

D.Garfincle, G.Horowitz  and  A.Strominger, {\it Phys.Rev.} {\bf D43},
3140 (1991), {\bf D45}, 3888 (1992).

\bibitem{c07}
J.Ellis, N.Kaloper,  K.A.Olive, J.Yokoyama, {\it Phys.Rev.} {\bf D59},
103509 (1999);

S.Mikohyama, {\it Phys.Rev.} {\bf D63}, 104025 (2001);

\bibitem{c02}
S.O. Alexeyev and  M.V. Pomazanov, {\it  Phys. Rev.} {\bf  D55},  2110
(1997);

S.O. Alexeyev and  M.V. Sazhin, {\it  Gen. Relativ. Grav.}  {\bf  30},
1187 (1998);

\bibitem{c03}
S.O. Alexeyev, M.V. Sazhin and M.V.Pomazanov, {\it Int. J. Mod. Phys.}
{\bf D10}, 225 (2001).

\bibitem{c04}
R.R. Metsaev, A.A. Tseytlin, {\it Phys. Lett.} {\bf B185}, 52 (1987);

\bibitem{levin}
J. J. Levin.
``Singular  perturbations   of   nonlinear   systems  of  differential
equations related to conditional stability''.
{\it Duke Mathematical Journal}, {\bf 23}, (1956).

\bibitem{tikhonov}
A. N. Tikhonov,
``Systems of differential equations containing small parameters in the
derivatives'',
{\it Math. Sbor. (in Russian)}, {\bf 31}, 576 (1952);

D. R. Smith,
``Singular-perturbation theory'',
{\it Cambridge University Press}, Cambridge, (1985).

\bibitem{c05}
Y.Kikuchi and C.Marzban, {\it Phys. Rev.} {\bf D35}, 1400 (1987).

\bibitem{c06}
Q-Han Park, D.Zanon, {\it Phys. Rev.} {\bf D35}, 4038 (1987).

\end{thebibliography}
\end{document}